\documentclass[aip,apl,reprint,superscriptaddress]{revtex4-1}
\usepackage{amsmath}
\usepackage{upgreek}
\usepackage{graphicx}
\usepackage{siunitx}
\usepackage{textcomp}
\newcommand{\tens}[1]{\overset{\leftrightarrow}{#1}}

\begin{document}
\setlength{\parskip}{0pt}
\title{Parallel Hall effect from 3D single-component metamaterials}

\author{Christian~Kern}
\affiliation{Institute of Applied Physics, Karlsruhe Institute of Technology (KIT), 76128 Karlsruhe, Germany}
\author{Muamer~Kadic}
\affiliation{Institute of Applied Physics, Karlsruhe Institute of Technology (KIT), 76128 Karlsruhe, Germany}
\author{Martin~Wegener}
\affiliation{Institute of Applied Physics, Karlsruhe Institute of Technology (KIT), 76128 Karlsruhe,
Germany}
\affiliation{Institute of Nanotechnology, Karlsruhe Institute of Technology (KIT), 76344 Eggenstein-Leopoldshafen, Germany}

\date{\today}

\begin{abstract}
We propose a class of three-dimensional metamaterial architectures composed of a single doped semiconductor (e.g., n-Si) in air or vacuum that lead to unusual effective behavior of the classical Hall effect. Using an anisotropic structure, we numerically demonstrate a Hall voltage that is parallel---rather than orthogonal---to the external static magnetic-field vector (``parallel Hall effect"). The sign of this parallel Hall voltage can be determined by a structure parameter. Together with the previously demonstrated positive or negative orthogonal Hall voltage, we demonstrate four different sign combinations.
\end{abstract}

\maketitle

In the textbook version of the classical Hall effect \cite{Yu10}, the Lorentz force due to an electric current in a semiconductor or metal plate and a magnetic-induction vector normal to the plate lead to an induced voltage. This Hall voltage is orthogonal to the current direction and orthogonal to the magnetic-field vector. In the linear regime, the Hall voltage is proportional to the magnetic field and proportional to the electric current. In isotropic bulk materials, the sign of the Hall voltage is determined by the dominant type of charged carrier, i.e., it is negative for n-doped semiconductors (or free-electron metals) and positive for p-doped semiconductors.

Recently, building upon the pioneering work of Briane and Milton \cite{Briane9}, we have shown \cite{Kadic15} that the sign of the Hall voltage can be modified by structure in a three-dimensional simple-cubic metamaterial lattice composed of intertwined tori made of only a single constituent material (e.g., n-doped silicon) in vacuum/air. We have argued that structures like that should be accessible experimentally by using current state-of-the-art three-dimensional fabrication technology. 
Due to the simple-cubic symmetry, the electrical properties have been isotropic \cite{Briane9, Kadic15}.

In this Letter, we show that an anisotropic single-component structure can lead to a Hall voltage parallel (rather than orthogonal) to the magnetic-field vector.
This finding complements the existing versions of the Hall effect, including the classical sign-inverted Hall effect \citep{Briane9, Kadic15}, the classical dynamic Hall effect \cite{Wegener2005}, the classical spin Hall effect \cite{SHE}, the quantum Hall effect \cite{QHE}, the fractional quantum Hall effect \cite{FQHE}, the quantum spin Hall effect\cite{QSHE}, and the optical spin Hall effect \cite{OSHE}. 

The mathematical treatment underlying this Letter has been described in Ref.\,\onlinecite{Kadic15} and shall thus be summarized only briefly here: Using the software package COMSOL Multiphysics (MUMPS solver, more than $10^4$ tetrahedral elements per unit cell and a relative tolerance of $10^{-4}$), we numerically solve the static continuity equation ${\vec{\nabla}\cdot\left(\tens{\sigma}  \vec{\nabla}\phi\right)=0}$ 
for the electrostatic potential $\phi=\phi(\vec{r})$ with the conductivity tensor $\tens{\sigma}=\tens{\sigma}(\vec{r})$. In the linear regime and for a magnetic-induction vector $\vec{B}=(0,0,B_z)$, it is given by
\begin{equation}
\tens{\sigma}= \sigma_0 \begin{pmatrix} 1 & \sigma_0 R_{\rm H} B_z & 0\\ -\sigma_0 R_{\rm H} B_z  & 1 & 0\\ 0 & 0 & 1\\ \end{pmatrix}\,.
\end{equation}
The current $I_x$ enters via the current density ${\vec{j}=\tens{\sigma}\vec{E}=-\tens{\sigma}\vec{\nabla}\phi}$ through the boundary condition 
${I_x=-\int_A \vec{j}\cdot\mathrm{d}\vec{A}}$, where the integration plane $A$ cuts the sample parallel to the $yz$-plane. To be concrete, we use material parameters corresponding to n-doped silicon as in \cite{Kadic15} with scalar conductivity $\sigma_0=200\,{\rm AV^{-1} m^{-1}}$ and Hall coefficient $R_{\rm H}=-624\cdot 10^{-6}\,{\rm m^3 A^{-1} s^{-1}}$, and $\tens{\sigma}=0$ for the air voids. However, our results can easily be scaled to other parameter sets. Our calculations are performed for a metamaterial Hall bar composed of a finite number of metamaterial unit cells $N_x$, $N_y$, and $N_z$ in the three spatial directions. Together with the cubic lattice constant $a$, this leads to the spatial extents of the Hall bar $L_x=aN_x$, $L_y=aN_y$, and $L_z=aN_z$. Large metal contacts (precisely, equipotential surfaces) parallel to the $yz$-plane covering the entire metamaterial faces are applied to impose the current $I_x$. For clarity, in this Letter, no metal contacts are explicitly added in the calculations to pick up the Hall voltages. We have previously shown \cite{Kadic15} that the corresponding perturbations are small. The shown Hall voltages are picked up at the unit cell in the center of the corresponding metamaterial sample surface.
\begin{figure}
	\includegraphics{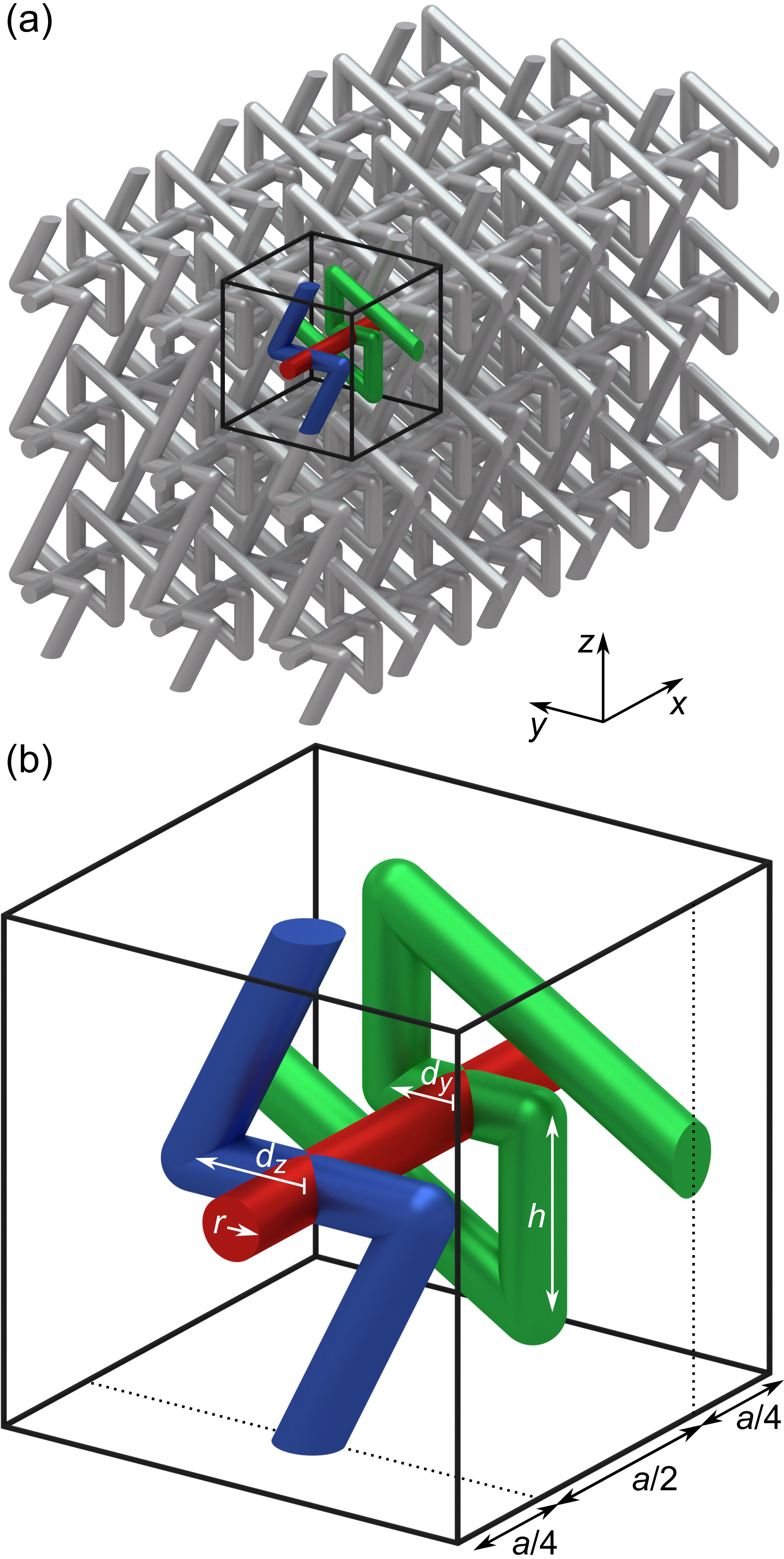}
    \caption{(a) Illustration of the suggested three-dimensional anisotropic metamaterial exhibiting a Hall voltage $U_z$ parallel to the magnetic-field vector $(0,0,B_z)$ in addition to the usual orthogonal Hall voltage $U_y$. The metamaterial is composed of a single constituent material, e.g., n-doped silicon, in air/vacuum. A constant current $I_x$ is imposed through the array of rods oriented along the $x$-direction. (b) One unit cell of the simple-cubic translational lattice with lattice constant $a$. The indicated parameters $d_y$ and $d_z$ can be negative/positive and determine the signs of $U_y$ and $U_z$, respectively. The three different functional parts are made of a single material, they are colored for illustration only. The current-carrying straight rods are highlighted in red. The structures highlighted in blue and green lead to Hall voltages in the $z$- and $y$-direction, respectively. The unit cell shown corresponds to a positive value of $d_z$ and a negative value of $d_y$, the latter leading to an inversion of the orthogonal Hall voltage.}
  \label{fig1}
\end{figure}
\begin{figure}
	\includegraphics{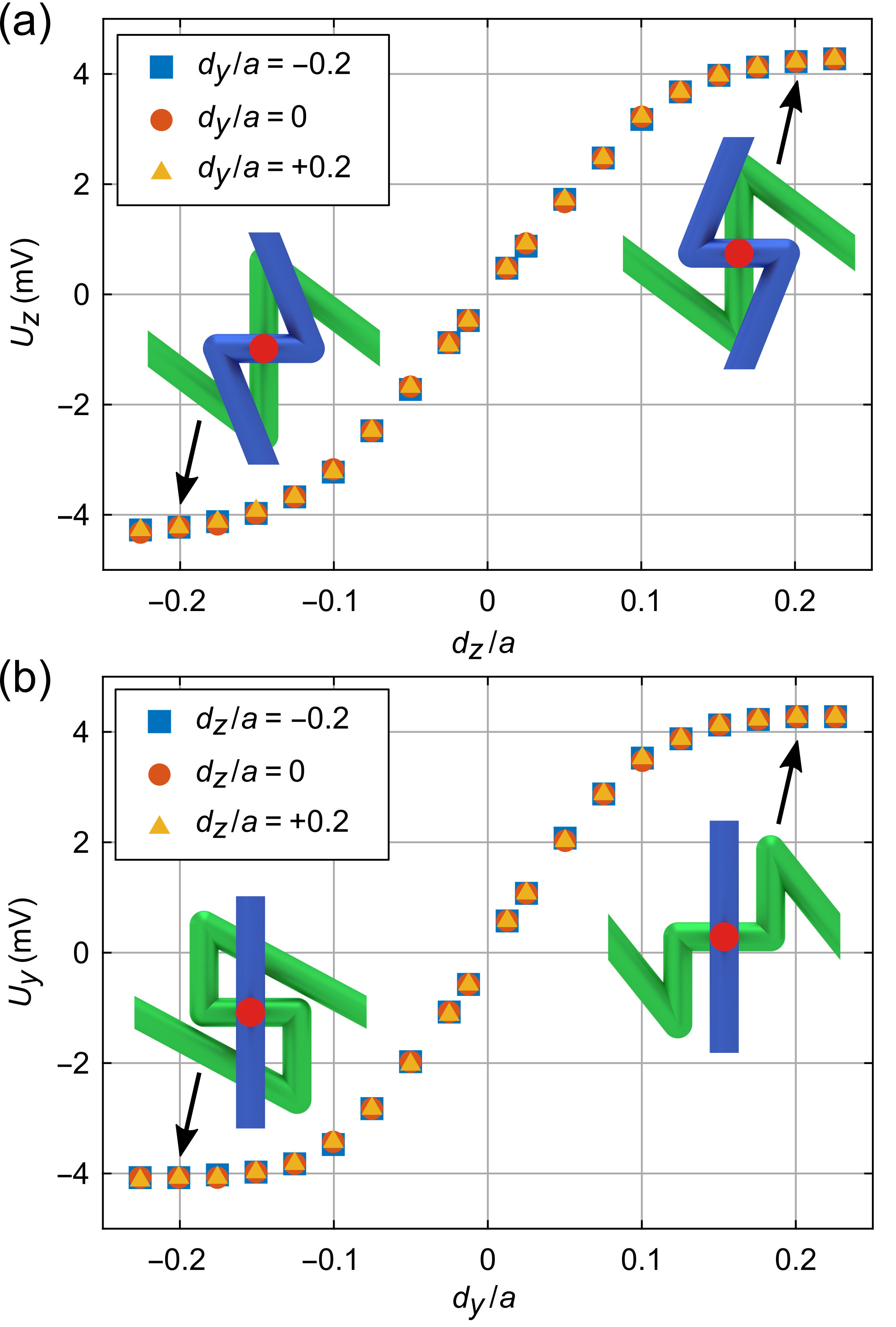}
    \caption{Calculated Hall voltages for the metamaterial structure illustrated in Fig.\,1. (a) $U_z$ versus $d_z$ for three different values of $d_y$. (b) $U_y$ versus $d_y$ for three different values of $d_z$. Parameters are: $\sigma_0=200\,{\rm AV^{-1} m^{-1}}$, $R_{\rm H}=-624\cdot 10^{-6}\,{\rm m^3 A^{-1} s^{-1}}$, $I_x=0.1\,{\rm mA}$, $B_z=1\,{\rm T}$, $N_x=9$, $N_y=N_z=3$, $h=15\,\si{\um}$, $r=2.5\,\si{\um}$ and $a=40\,\si{\um}$.}
  \label{fig2}
\end{figure}

For the static case discussed in this Letter, the electromagnetic wavelength, $\lambda$, is formally infinitely large and, hence, an effective-medium description requiring the condition $\lambda/a \gg 1$ to be fulfilled is meaningful for any lattice constant $a$, e.g., for micrometer and centimeter scales. However, macroscopic structures are less favorable because the Hall voltage is proportional to the Hall resistance, which, like any resistance, scales inversely proportional to size (if all dimensions are scaled simultaneously). This means that, to obtain absolute Hall voltages that are readily measurable, macroscopic structures require orders of magnitude larger absolute currents $I_x$ and/or larger magnetic fields $B_z$. Furthermore, the magnetic fields obviously need to be homogeneous over much

larger regions in space for macroscopic structures. We have thus chosen a lattice constant of $a=40\,\si{\um}$ in all of the calculations shown in this Letter. We believe that the corresponding three-dimensional structures can be fabricated by using three-dimensional direct laser writing (as discussed in more detail in Ref.\,\onlinecite{Kadic15}).

Intuitively speaking, the sign of a Hall voltage can simply be flipped by interchanging the two wires used to pick up this Hall voltage from the Hall bar. 
To avoid short circuits due to crossings of wires, this interchanging requires to go out of a common plane, i.e., it requires a non-planar three-dimensional geometry \cite{Briane9}. In the metamaterials previously suggested in Ref.\,\onlinecite{Kadic15}, this interchanging has rather been accomplished locally inside of the metamaterial by intertwined tori within each three-dimensional metamaterial unit cell. 
\begin{figure*}
	\includegraphics{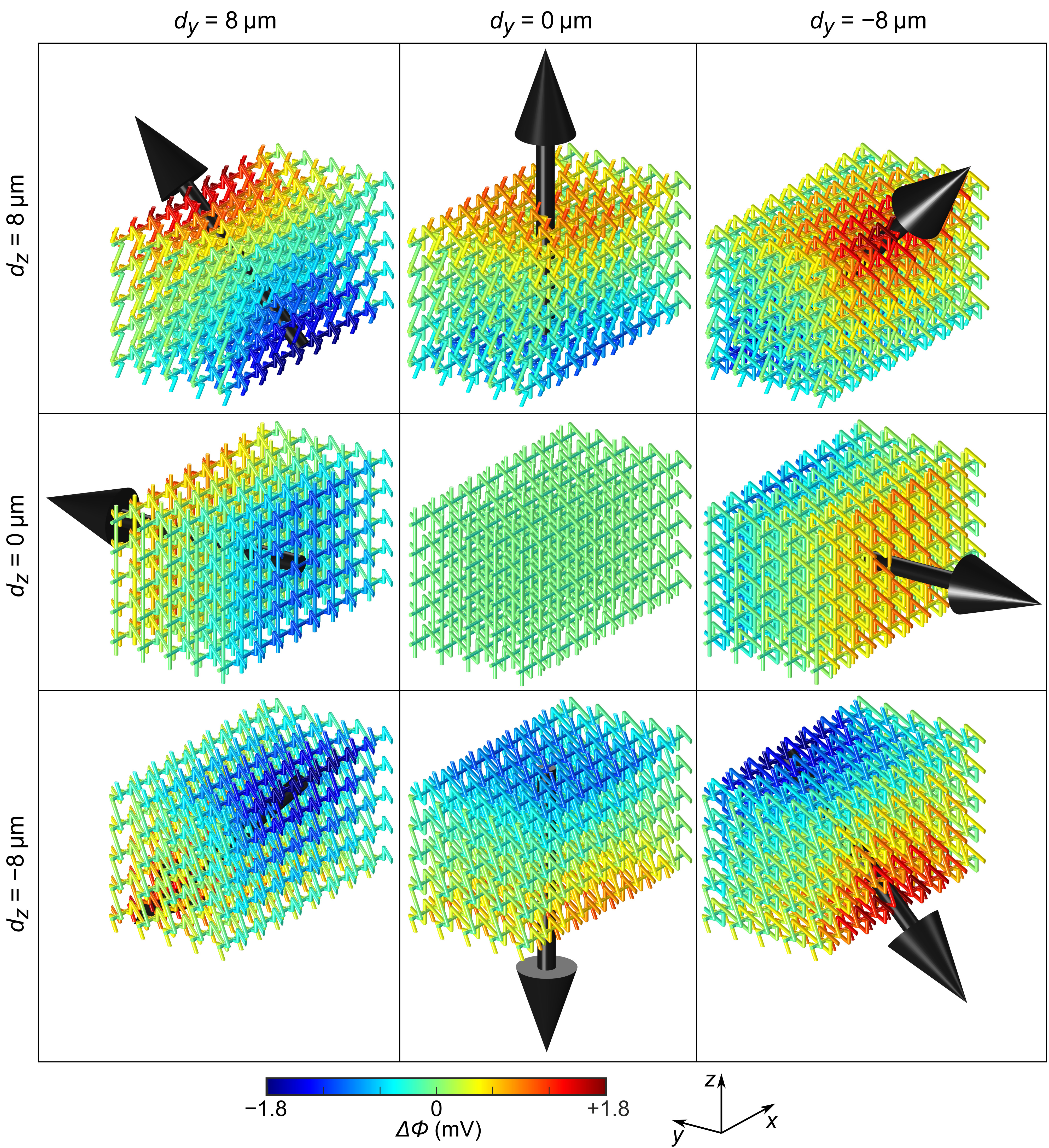}
    \caption{Nine examples of Hall potential maps, i.e., electrostatic potential difference $\mathit{\Delta}\phi(\vec{r})=\phi(\vec{r},B_z=1\,{\rm T})-\phi(\vec{r},B_z=0)$ for $N_x = 9$, $N_y = N_z = 5$ and different values of $d_y$ (different columns) and $d_z$ (different rows), plotted on a common false-color scale. The parameters are the same as in Fig.\,2. The black arrows indicate the direction of the Hall-potential gradient.}
  \label{fig3}
\end{figure*}

Expanding along this line, the novel blueprint depicted in Fig.\,1 uses an array of straight semiconductor rods oriented along the $x$-direction (red in panel (b)). 
A total electrical current $I_x$ flows via these rods through the entire sample. As a result, one locally gets an ordinary Hall voltage across each of these wires along the $y$-direction. The structure contains two additional sets of ``pick-up wires'' made of the same material. One set of wires (green) connects the individual Hall voltages, which can be seen as local voltage sources, and adds them all up along the $y$-direction, leading to the usual orthogonal Hall voltage $U_y$. Likewise, the local voltages add up along the $z$-direction through the second set of wires (blue), leading to the voltage $U_z$ parallel to the magnetic-field direction. The two crucial geometry parameters are the quantities $d_y$ and $d_z$ 
(see Fig.\,1). If they are negative, the sign of each local voltage source is reversed and so are the macroscopic voltages $U_y$ and $U_z$, respectively. The results of our numerical calculations shown in Fig.\,2 clearly confirm this conjecture. 

An illustrative way to see what is going on inside of the Hall bar is to inspect the Hall potential. We define the Hall potential as the difference of the electrostatic potential with magnetic field and that without. In this fashion, we subtract the overwhelmingly large but trivial potential ramp due to the applied bias voltage leading to the imposed current $I_x$. The examples of the Hall-potential maps depicted in Fig.\,3 demonstrate that the Hall voltages always build up from the ``bulk'' as opposed to being a mere surface effect. Furthermore, the data in Fig.\,3 show that we gain control of the orientation of the Hall-potential gradient within the $yz$-plane (see black arrows). Equivalently, we can say that we have achieved unusual effective metamaterial conductivity tensors $\tens{\sigma}$ (see above) with adustable non-zero elements $\sigma_{xz}=-\sigma_{zx}$ and $\sigma_{xy}=-\sigma_{yx}$.

As usual, the size and the aspect ratios of the Hall bar play an important role \cite{Moelter98} regarding whether one observes ``bulk'' effects or whether edge effects dominate. Figure\,4 illustrates the scaling of the parallel Hall voltage $U_z$ versus the length-to-height ratio of the Hall bar (i.e., versus $N_x/N_z$). The depicted results resemble the typical behavior of the geometrical correction factor for ordinary bulk materials \cite{Moelter98}. Additionally, it becomes clear that considering Hall bars composed of merely $9\times 3\times 3$ unit cells (as used for the calculations depicted in Fig.\,2) is already sufficient to approach convergence towards case of infinitely many unit cells.

For such sufficiently large and suitably shaped samples \cite{Moelter98} and constant current $I_x$, the ordinary orthogonal Hall voltage ($U_{\rm H}=U_y$) is proportional to $1/L_z$ and neither depends on $L_x$ nor on $L_y$. In contrast, the parallel Hall voltage $U_z$ is proportional to $1/L_y$ and does not depend on $L_x$ and $L_z$. Here, the increase of the Hall voltage with the number of unit cells along the $z$-direction is compensated by the fact that the current through one unit cell scales inversely proportional to the number of unit cells along $z$. The current through one unit cell and hence $U_z$ also scales inversely proportional to the number of unit cells along the $y$-direction. We have numerically verified all of these scaling laws (not depicted).
\begin{figure}
	\includegraphics{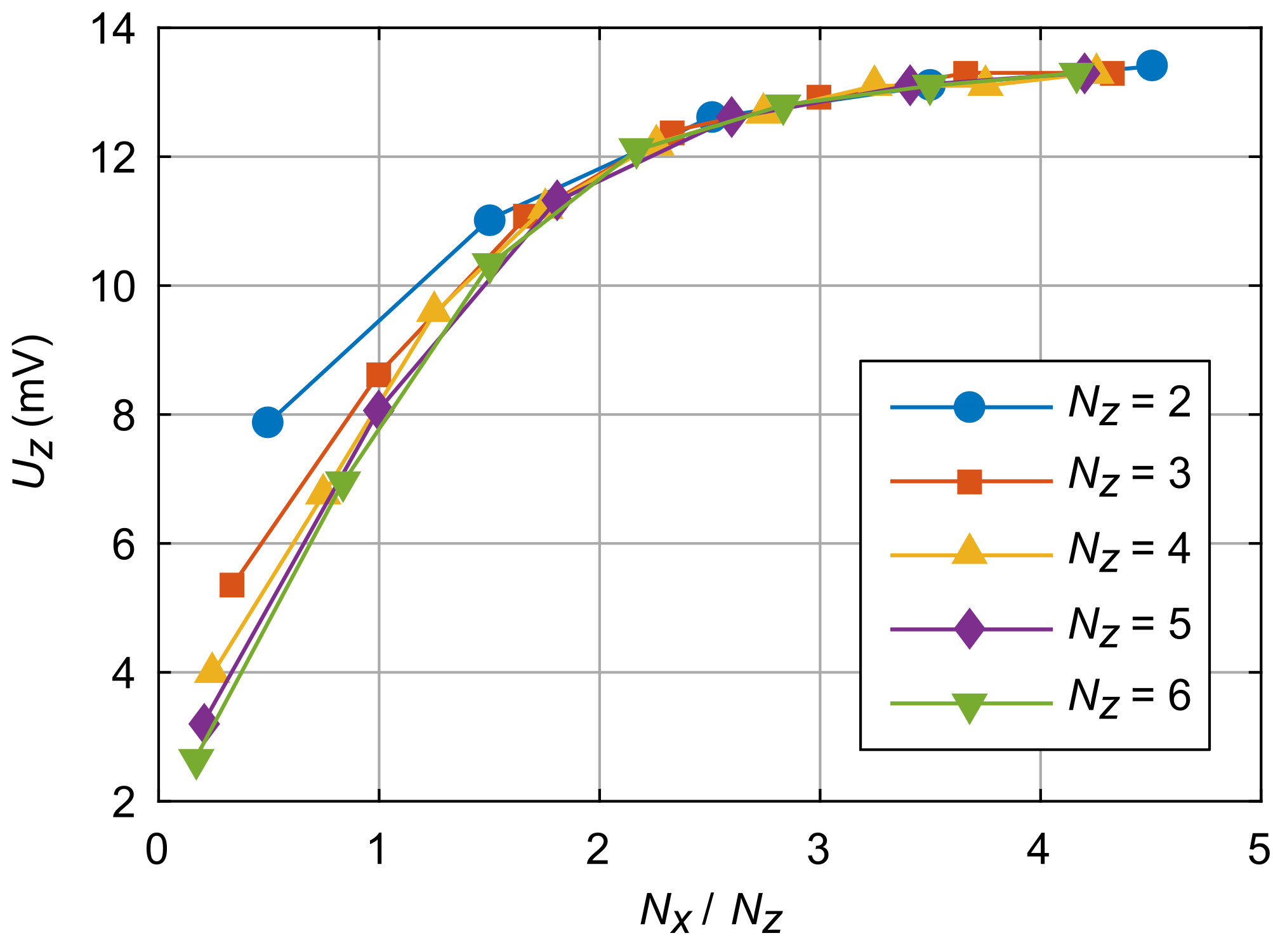}
    \caption{Scaling of the parallel Hall voltage versus the length-to-height ratio $N_x/N_z$. Parameters are $N_y=1$, $d_y=-8\,\si{\um}$ and $d_z=10\,\si{\um}$. The other parameters are as in Fig.\,2.}
  \label{fig4}
\end{figure}

In the special case of $d_y=0$ and $d_z=h$ (see Fig.\,1), we obtain a metamaterial exhibiting the parallel Hall effect for an arbitrary orientation of the magnetic-field vector in the $yz$-plane (not depicted). In this special case, the class of structures we suggest shares some similarities to the structure suggested earlier by Briane and Milton \cite{Briane10}. However, their structure requires two different constituent materials, one of which has anisotropic electrical properties. In sharp contrast, our structure is composed of only a single locally electrically isotropic constituent material (like, e.g., n-doped silicon) -- which is much simpler to realize experimentally. Furthermore, it is conceptually even more striking to obtain the parallel Hall effect using only a porous structure of a single bulk constituent material. Finally, we note that the qualitative results described in this Letter can also be obtained by using hollow rods instead of massive rods (not depicted). This fact enlarges the possibilities for experimental realizations. 

In conclusion, we have suggested three-dimensional metamaterials effectively exhibiting the parallel Hall effect. These metamaterials are composed of only a single constituent material in vacuum/air, i.e., they can be considered as a porous version of a bulk material. Broadly speaking, our results exemplify that the effective properties of metamaterials can be way off those of the constituent(s).

We acknowledge support by the Hector Fellow Academy and by the Karlsruhe School of Optics \& Photonics (KSOP).

\end{document}